\documentclass[a4paper,10pt]{article}
\usepackage[utf8x]{inputenc}

\usepackage{amsmath,amssymb,bm}
\usepackage{sidecap}
\usepackage{times}
\usepackage{graphicx}
\usepackage{float}
\usepackage[left=2.5cm,top=2.5cm,right=2.5cm,bottom=2.5cm,nohead,nofoot]{geometry}

\date{}

\begin{document}

\begin{centering}
\LARGE\textbf{Symmetry dependent electron localization and optical absorption of polygonal quantum rings}
       
\vspace{12pt}      
\normalsize\textbf{Anna Sitek$^{\ast,\ddagger}$, Vidar Gudmundsson$^{\ast}$, and Andrei Manolescu$^{\P}$}

\vspace{0pt}  

\normalsize\textit{$^{\ast}$Science Institute, University of Iceland, Dunhaga 3, IS-107 Reykjavik, Iceland}\\
\normalsize\textit{$^{\ddagger}$Department of Theoretical Physics, Wroclaw University of Technology, 50-370 Wroclaw, Poland}\\
\normalsize\textit{$^{\P}$School of Science and Engineering, Reykjavik University, Menntavegur 1, IS-101 Reykjavik, Iceland}\\
\normalsize\textit{e-mail: sitek@hi.is}\\
\end{centering}
\vspace{6pt}

\noindent
\textbf{ABSTRACT}\\
We compare energy spectra, electron localization and optical absorption of 
square and diamond quantum rings and analyze how sample geometry affects
those features. We show that low energy levels of diamond rings form two 
groups delocalized between opposite corners which results in increased 
number of optical transitions. We also show that contacts applied to corner 
areas allow for continuous change between square- and diamond-like behavior 
of the same sample, irrespective of its shape.


\vspace{12pt}
\noindent
\textbf{1. INTRODUCTION}
\vspace{3pt}

\noindent
Polygonal quantum rings are very short hollow or core-multi-shell wires in 
which electrons are confined only in one of the shells. The most common are 
hexagonal structures, but dodecagonal \cite{Rieger15}, 
triangular \cite{Qian04,Qian05} and tetragonal \cite{Kavanagh12} core-shell 
wires have already been achieved. 
The main feature which distinguishes polygonal from circular structures is 
a unique carrier localization, which in the case of quantum wires leads to 
a formation of one-dimensional channels 
\cite{Jadczak14, Bertoni11, Royo13, Royo14, Fickenscher13, Shi15}.
As in the case of bent quantum wires \cite{Sprung}, in the corner areas of 
polygonal rings effective quantum wells are formed which attract low-energy 
electrons and localize them only in the vicinity of the vertices  
\cite{Ballester12, Sitek15}. If the rings are externally and internally 
restricted by regular polygons, then the localization probability associated 
with the lowest states is equally distributed between all corners, but this
may be easily turned into localization in single corners if the system 
symmetry is broken. Moreover, the existence of corners changes the energy 
degeneracy, i.e. splits degeneracies related to angular momentum conservation. 
For sufficiently narrow rings the polygon separates energy levels for 
which the probability distribution is localized only in the corner areas from 
higher, mostly side-localized, eigenvalues. Electron distribution also affects 
absorption of electromagnetic waves. In the presence of an external magnetic 
field and circularly polarized light only two transitions, each coupled to 
a different polarization, from the ground state to the corner- or side-localized 
states above the energy gap occur for symmetric samples \cite{Sitek15}.

In this paper we focus on square and diamond shaped rings, we show how 
probability distribution changes when the system symmetry is reduced, and how 
it affects absorption of electromagnetic field. We compare two cases: In the 
first case the system geometry is variable; we show that for a diamond ring 
the energy levels are only spin degenerated and the corner-localized states 
form two energy groups. The lower one with probability distribution equally 
shared by the two sharper corners and the higher group associated with 
electrons bound symmetrically to the wider corners. We show that diamond 
samples allow for twice as many optical transition than the square rings in 
the presence of one polarization type. In the second case we apply potentials 
to the corner areas and show how they allow to switch between square- and 
diamond-like carrier localization and absorption for one or another ring shape.

\vspace{12pt}
\noindent
\textbf{2. THE MODEL}
\vspace{3pt}

%--------------------------------------------------------------------
\begin{SCfigure}[5][h]
\centering
\includegraphics[width=0.21\textwidth,angle=0]{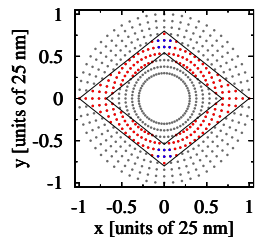}
 \caption{\textit{Sample model - diamond constraints, for with the ratio between 
          the diagonals is $0.8$, applied on polar grid. Blue points 
          indicate areas where an on-site potential was applied. 
          For visibility we reduced the number of site points.}}
 \label{Fig:Sample}
\end{SCfigure}

%-------------------------------------------------------------------- 

% %--------------------------------------------------------------------
% \begin{figure}[H]
% \centering
% \includegraphics[width=0.21\textwidth,angle=0]{Sample_D.eps}
% %  \includegraphics[scale=0.67]{Sample_QR.eps}
%  \caption{\textit{Sample model - diamond constraints, for with the ratio between 
%           the diagonals is $0.8$, applied on polar grid. Blue points 
%           indicate areas where an on-site potential was applied. 
%           For visibility we reduced the number of site points.}}
%  \label{Fig:Sample}
% \end{figure}
% 
% %-------------------------------------------------------------------- 

\noindent
The sample model used in our calculations is based on a discretization method 
on a polar grid with diamond constraints defining the ring shape. The Hilbert 
space is  spanned by vectors $|kj\sigma\rangle$, where $k$ and $j$ refer to 
the radial and angular coordinates, respectively, associated only with sites 
within the polygonal shell, $\delta r$ and $\delta \phi$ are the corresponding 
intervals and $\sigma$ denotes spin. The Hamiltonian matrix element in 
these coordinates is \cite{Daday11} 
%----------------------
\begin{eqnarray}
\label{Hamiltonian}
H_{kj\sigma,k'j'\sigma'} = 
 T\delta_{\sigma,\sigma'}\!\!\!\!\!\!\!\!\!\!\! & &\left[\left( t_r +  t_{\phi} 
 + \frac{1}{2}t^2_{\mathrm{B}}\left(\frac{r_k}{4R_{\mathrm{ext}}}\right)^2 + V\right)
 \delta_{k,k'} \delta_{j,j'}\right.\\ \nonumber
 & & \left. -\left(t_{\phi} + t_{\mathrm{B}}\frac{i}{4\delta\phi}\right)\delta_{k,k'} \delta_{j,j'+1} 
 + t_{r}\delta_{k,k'+1} \delta_{j,j'} + \mathrm{H.c.}\right]
  + \frac{1}{2}Tt_{\mathrm{B}}\gamma\left(\sigma_z\right)_{\sigma,\sigma'}\delta_{k,k'}\delta_{j,j'},
\end{eqnarray}
%----------------------
where $T=\hbar^{2}/(2m^{*}R^{2}_{\mathrm{ext}})$ is an energy factor, $m^{*}$ 
the effective mass of the semiconductor material, $R_{\mathrm{ext}}$  is the 
external radius of the polar grid, $t_r=(R_{\mathrm{ext}}/\delta r)^2$, 
$t_{\phi}=[R_{\mathrm{ext}}/(r_k\delta\phi)]^2$, $t_{\mathrm{B}}=\hbar eB/(m^{*}T)$ 
is the cyclotron energy in units of $T$, $e$ is the electron charge and 
$B$ a magnetic field perpendicular to the ring plane (which lifts energy 
degeneracies due to spin and angular momentum), $\sigma_{z}$ stands for 
the $z$th Pauli matrix, $\gamma=g^{*}m^{*}/(2m_{e})$ is the ratio between the 
Zeeman gap and the cyclotron energy with $g^{*}$ being the electron $g$-factor, 
$m_e$ the free electron mass, and $V$ stands for an on-site potential.

The optical absorption coefficient is calculated in the dipole and zero 
temperature approximations according to Refs.\ \cite{Haug09,Chuang95,Hu00}
and equals 
%----------------------
\begin{eqnarray*}
\label{absorption_coeff}
 \alpha(\hbar\omega) = A\hbar\omega\sum_{\mathrm{f}}|\langle f|\bm{\varepsilon}\cdot\bm{d}|i\rangle|^2
 \delta\left(\hbar\omega - \left(E_{\mathrm{f}}-E_{\mathrm{i}}\right)\right),
\end{eqnarray*}
%----------------------
where  $A$ is a constant, $\bm{\varepsilon}=\left(1,\pm i\right)/\sqrt{2}$ 
the circular polarization of the electromagnetic field, $\bm{d}$ the dipole 
moment and $E_{\mathrm{i,f}}$ the energies of the initial and final states 
$|i,f\rangle$, respectively. The delta function was approximated by a 
Lorentzian $\left(\Gamma/2\right)/\{\left[\hbar\omega
-\left(E_{\mathrm{f}}-E_{\mathrm{i}}\right)\right]^2
+ \left(\Gamma/2\right)^2\}$,
where $\Gamma$ is a phenomenological broadening. Since we neglect spin-orbit 
coupling, optical transitions do not allow for spin flip.

\vspace{12pt}
\noindent
\textbf{3. RESULTS}
\vspace{3pt}

\noindent
In all of the cases analyzed below the external radius of the disk-shaped
grid, which is also the largest radius of the tetragons, is set equal 
to $25$ nm and the side thicknesses are equal to $5$ nm. The material
parameters correspond to InAs, where 
$m^{*} = 0.023m_{e}$ and $g^{∗}=-14.9$ and thus the energy unit $T$ 
introduced in the Hamiltonian (\ref{Hamiltonian}) is approximately 
$2.8$ meV and the ratio $\gamma = −0.171$. The samples consist of 
over $6000$ grid points.

%--------------------------------------------------------------------
\begin{figure}[h]
\centering
\includegraphics[scale=0.66]{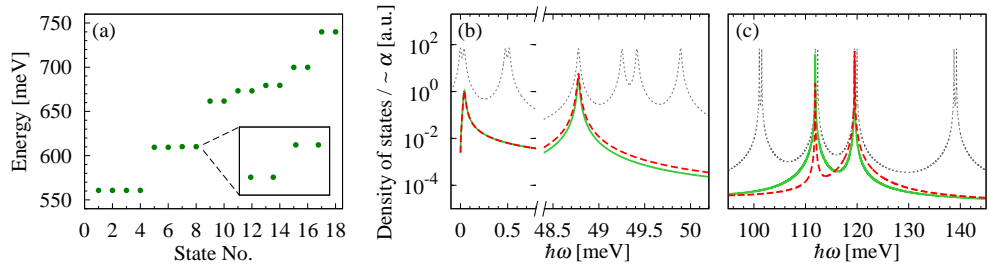}
\caption{\textit{Energy levels (a). Density of states (gray - dotted) 
         and absorption coefficients in the presence of clockwise 
         (red - dashed) and counterclockwise (green - solid) 
         polarization for the diamond ring shown in Fig.\ \ref{Fig:Sample} 
         initially containing one electron in the ground state [(b) and (c)].}}
\label{fig:diamond_e_abs}
\end{figure}
%--------------------------------------------------------------------

%--------------------------------------------------------------------
\begin{figure}[h]
\centering
\includegraphics[scale=0.66]{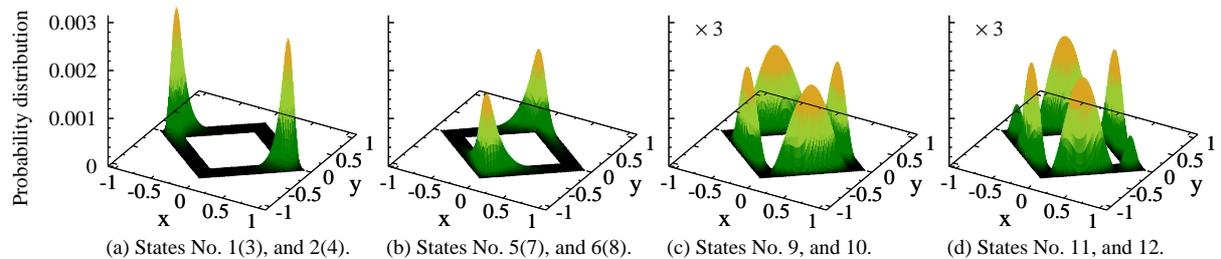}
\caption{\textit{Probability densities corresponding to the first 12 levels 
         shown in Fig.\ \ref{fig:diamond_e_abs}(a).  Corner states of the lowest 
         (a) and second (b) energy groups, then purely side localized states (c), 
         and finally higher states where side and 
         corner localization coexists (d). For proper comparison the probability
         density in (c) and (d) was scaled by a factor of three.}} 
\label{fig:localization_diamond}
\end{figure}
%--------------------------------------------------------------------

In a recent study we analyzed regular polygons \cite{Sitek15}. In the 
present paper we discuss what happens when the symmetry of the polygon 
is changing. To do so we modify the square ring by decreasing one diagonal
to $40$ nm and obtain a diamond quantum ring as shown in 
Fig.\ \ref{Fig:Sample}. Several low energy levels of this sample are 
plotted in Fig. \ref{fig:diamond_e_abs}(a), where one can distinguish 
two groups of states which look like four-fold degenerated energy levels, 
but in fact they are composed of two close-by spin (two-fold)  degenerated 
eigenvalues [inset to Fig.\ \ref{fig:diamond_e_abs}(a)]. The shape and 
depth of the effective wells formed in the vicinity of the vertices depend 
on corner angles and areas. Sharper corners have a larger area between the 
external and the internal boundaries of the polygon, which results in
formation of deeper wells. Thus the lowest energy states, No. 1-4, are
localized in the sharpest corners [\ref{fig:localization_diamond}(a)], 
and the higher energy states, No. 5-8, are spread between the (shallower) 
wells existing in the wider corners [\ref{fig:localization_diamond}(b)].
This results in formation of an energy gap of about 50 meV in the corner
state domain.  For a square polygon the two groups of four states merge 
into a single group of eight states with a dispersion of about 5 meV 
\cite{Sitek15} (also shown below).
For the higher levels of the energy spectrum the probability distributions 
are spread over the polygon sides. The distribution corresponding to the 
first energy level above the corner localized groups, states No. 9-10, 
is purely delocalized between the side areas 
[Fig.\ \ref{fig:localization_diamond}(c)] {similarly to the side states of 
a square ring.  The localization pattern of the higher states includes 
a small probability of finding the electron in corner areas, but here in 
the sharper corners for the second state above the corner-localized group, 
No. 11-12 [Fig.\ \ref{fig:localization_diamond}(d)] or in the wider corner 
areas for the next energy level (not shown). This means that electron 
distribution for a diamond ring differs considerably from the one of a 
square sample only in the low-energy domain.

The symmetry reduction and formation of two different wells affects
absorption of electromagnetic field. To remove the spin degeneracy
we consider the diamond sample immersed in a weak magnetic field
of $0.53$ T, perpendicular to its plane, which produces a spin
splitting of 0.48 meV.  We assume the sample initially contains an
electron in the ground state.  As seen in Fig.\ \ref{fig:diamond_e_abs},
light which is circularly polarized in the sample plane may excite the
electron to two other corner states [Fig.\ \ref{fig:diamond_e_abs}(b)]
as well as to side-localized states [Fig.\ \ref{fig:diamond_e_abs}(c)] 
and all four transitions occur in the presence of both clockwise and
counterclockwise polarization types. In the analogous case of 
a square ring only two, complementary transitions occur for each 
type of polarization \cite{Sitek15}.

%--------------------------------------------------------------------
\begin{figure}[H]
\centering
\includegraphics[scale=0.66]{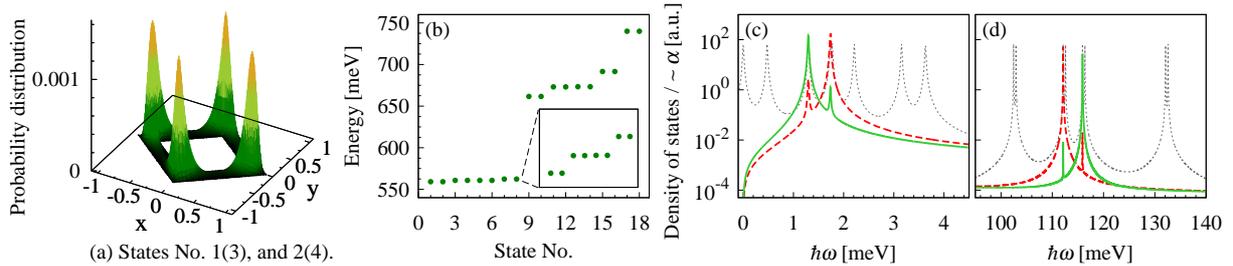}
\caption{\textit{Probability distribution associated with the ground state 
         of a diamond sample with an on-site potential equal to $372$ meV (a), 
         corresponding energy levels (b), and absorption spectrum for magnetic field 
         equal to $0.53$ T [(c) and (d)].}}
\label{fig:on_site}
\end{figure}
%--------------------------------------------------------------------

An external electric field allows to control electron distribution within
a ring. If it is applied in the plane of the polygon it affects all of 
the corners and changes the geometry of the effective wells formed
in their vicinity. This may easily break wave function symmetry, and
thus open all spin allowed transitions, or partially restore symmetric
electron localization in asymmetric samples \cite{Sitek15}. Even more
precise control may be achieved if point contacts are applied in the
corner areas which enable to control the depth of each well separately.

To model this later situation we applied an on-site potential in the
wider corner areas of diamond sample as indicated by the blue points in
Fig.\ \ref{Fig:Sample}. Negative potential deepens the quantum wells and 
thus shifts the energy levels of the sample and delocalizes low-energy 
carriers between all of the corners. The ratio of probability maxima 
depends on the value of the external potential and may be continuously 
adjusted, i.e., probability distribution may become equally distributed 
between all four corners as in the case of an ideal square ring  
[Fig.\ \ref{fig:on_site}(a)]. In this case the energy spectrum is nearly 
indistinguishable from the one of square sample [Fig.\ \ref{fig:on_site}(b)
and Ref.  \cite{Sitek15}], but even levels which seem to be four-fold 
degenerated consists of pairs of very close only spin degenerated 
eigenvalues. In the case shown in Fig.\ \ref{fig:on_site} this splitting 
is a results of a small mismatch of the potential and could be reduced 
even further. As long as all corners are populated, even if the 
localization peaks differ considerably from each other, the energy spectrum 
resembles more the one of a square sample \cite{Sitek15} than of a diamond 
ring [Fig.\ \ref{fig:diamond_e_abs}(a)].

Interesting effects are observed in the absorption spectrum if all of the 
corners are nearly equally populated. As seen in 
Figs.\ \ref{fig:on_site}(c) and \ \ref{fig:on_site}(d), two 
transitions from the ground state to corner- and side-localized states 
occur, but in each domain one of the absorption coefficient's maximum is 
about $1000$ times smaller from the other one. Thus, irrespective of the 
sample shape, only two relevant transitions associated with one polarization 
type take place, as in the case of square rings \cite{Sitek15}.

\vspace{12pt}
\noindent
\textbf{4. CONCLUSIONS}
\vspace{3pt}

\noindent
We studied electron localization and optical absorption of a diamond
quantum ring and compared it to the square sample. We showed that this
geometry induces two groups of low-energy states localized either in the
sharper or in the wider corners, respectively. The probability 
distributions associated with higher energy levels are mostly spread 
over all polygon sides and resemble those of square rings. Diamond rings 
allow for more optical transitions in the presence of one circular 
polarization type than square rings. We also showed that gates applied 
in corner areas enable to achieve a wide range of electron localization 
patterns and thus different energy and absorption spectra within one 
sample irrespective of its shape which opens wide range of control 
possibilities.

\vspace{12pt}
\noindent
\textbf{ACKNOWLEDGMENTS}
\vspace{3pt}

\noindent
This work was financially supported by the Research Fund of the University 
of Iceland, and the Icelandic Instruments Fund. We also acknowledge support 
from the computational facilities of the Nordic High Performance 
Computing (NHPC), and the Nordic network NANOCONTROL.

\bibliographystyle{plain}

\end{document}